\documentclass[cameraready]{Interspeech}

\title{Noise-Aware In-Context Learning for Hallucination Mitigation in ALLMs}

\author[affiliation={}]{Qixuan}{HUANG}
\author[affiliation={}]{Khalid}{Zaman}
\author[affiliation={}]{Masashi}{UNOKI}

\address{
     Graduate School of Advanced Science and Technology, \\
   Japan Advanced Institute of Science and Technology \\
   1-1 Asahidai, Nomi, Ishikawa 923--1292 Japan
}

\email{\{qixuan, zaman.khalid, unoki\}@jaist.ac.jp}

\keywords{Auditory large language model, Auditory hallucination, Benchmark, In-context learning}

\usepackage{comment}

\begin{document}

\maketitle

\makeatletter
\renewcommand{\footnoterule}{}
\makeatother

\begin{abstract}

Auditory large language models (ALLMs) have demonstrated strong general capabilities in audio understanding and reasoning tasks. However, their reliability is still undermined by hallucination issues. Existing hallucination evaluation methods are formulated as binary classification tasks, which are insufficient to characterize the more complex hallucination patterns that arise in generative tasks. Moreover, current hallucination mitigation strategies rely on fine-tuning, resulting in high computational costs. To address the above limitations, we propose a plug-and-play Noise-Aware In-Context Learning (NAICL) method. Specifically, we construct a noise prior library, retrieve noise examples relevant to the input audio, and incorporate them as contextual priors, thereby guiding the model to reduce speculative associations when acoustic evidence is insufficient and to adopt a more conservative generation strategy. In addition, we establish a hallucination benchmark for audio caption tasks including the construction of the Clotho-1K multi-event benchmark dataset, the definition of four types of auditory hallucinations, and the introduction of metrics such as hallucination type distribution to support fine-grained analysis. Experimental results show that all evaluated ALLMs exhibit same hallucination behaviors. Moreover, the proposed NAICL method reduces the overall hallucination rate from 26.53\% to 16.98\%.

 
\end{abstract}

\begingroup
\renewcommand\thefootnote{}        
\renewcommand\footnoterule{}       
\footnotetext{
This research was supported by a research grant from
\ifcameraready
JSPS KAKENHI (25H01139).
\else
an anonymous anonymous anonymous.
\fi
}%
\addtocounter{footnote}{-1}        
\endgroup

\section{Introduction}

Auditory large language models (ALLMs) have demonstrated strong general capabilities in audio understanding and reasoning \cite{Wang2025AudioBench,qian2025perception}. However, in real world audio scenarios characterized by overlapping events, background noise, and pervasive acoustic–semantic uncertainty, models tend to rely on linguistic priors to produce overly deterministic interpretations, thereby giving rise to auditory hallucinations \cite{ma2025towards}.

Existing studies have primarily evaluated ALLMs through classification tasks \cite{kuan2024understanding,kuan2025can}. Kuan et al. \cite{kuan2024understanding} proposed an evaluation framework for object hallucination, in which a large number of hallucination questions with “No” as the correct answer are constructed via negative sampling under a discriminative setting, and classification metrics are used to analyze model biases and error patterns in judging the existence of sound sources or events. They introduced a generative captioning task, extracting nouns from generated captions as an object set to measure the proportion of objects that are mentioned but do not actually exist. Subsequently, Kuan and Lee \cite{kuan2025can} extended discriminative evaluation to object existence, temporal order, and object attributes to assess models’ understanding of event changes and temporal relationships. However, classification evaluation methods tend to conflate the omission of genuinely present events with hallucinations, and the focus of existing generative evaluations is limited to the object-level. Existing audio datasets also exhibit limitations for hallucination research. Although AudioSet \cite{gemmeke2017audio} constructs a unified ontology covering huge number of sound events, its annotation scheme is biased toward small events and includes some non-real-world audio such as game clips. Clotho \cite{drossos2020clotho}, while providing multiple descriptions to capture semantic diversity, is still affected by annotators’ subjective focus differences, which can lead to inconsistencies across descriptions, and similarly offers limited coverage of subtle or marginal events \cite{morato2021diversity}.



To address these issues, we proposes a Noise-Aware In-Context Learning (NAICL) method for mitigating auditory hallucinations and construct a benchmark for fine-grained auditory hallucination analysis. For hallucination mitigation, instead of treating noise as simple interference or data augmentation, our method models diverse broadband noise as a class of acoustic priors with weak semantic properties and retrieve them during inference as contrastive contextual examples. This noise-prior contrastive mechanism guides the model to reduce the degree of semantic commitment when acoustic evidence is insufficient, avoiding over-completion driven by linguistic priors and thereby effectively suppressing hallucinations in generative audio understanding.

The benchmark construction in this study is conducted based on the Clotho dataset.
First, samples with insufficient acoustic evidence are removed, as even human listeners cannot reach consistent judgments on them. The original annotations are then manually revised to remove content inconsistent with the acoustic evidence and to supplement perceptually reliable events that were previously omitted.
As a result, a high-quality reference dataset of multi-event audio samples is constructed. In the evaluation stage, deviations between model-generated outputs and the reference descriptions are analyzed. In evaluation process, we utilize an LLM as the evaluator to compare generated outputs with the manually revised references. Hallucinations are then categorized into four types with clearly defined semantic boundaries: prior-driven, fabricated events, source or material, and acoustic attribute.


In summary, this paper presents an evaluation benchmark and mitigation method for hallucinations in ALLMs, with the following main contributions:

\begin{figure}[t]
  \centering
  \includegraphics[width=\linewidth]{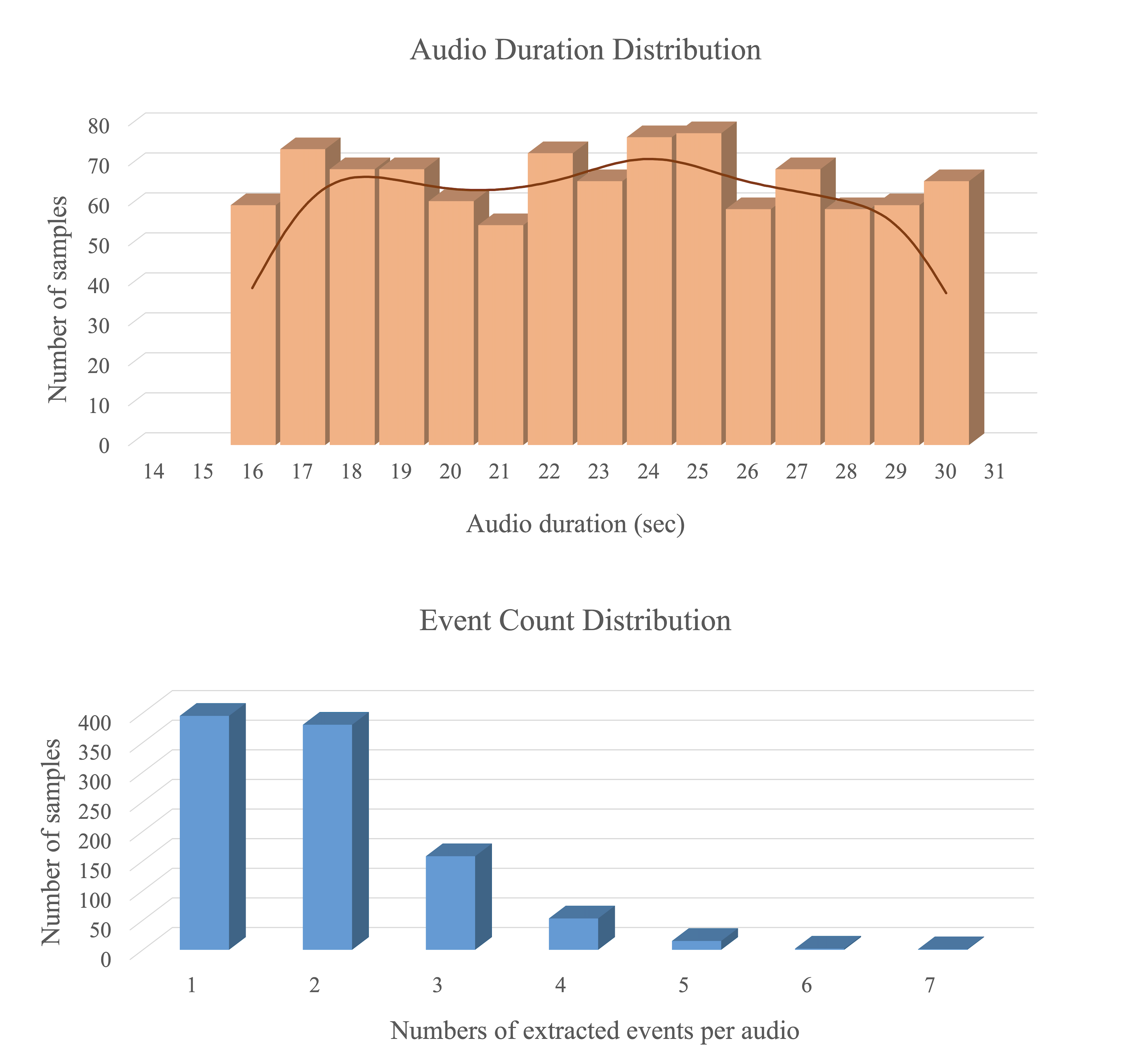}
  \caption{Distribution of audio durations in Clotho-1K benchmark and event counts based on AudioSet ontology.}
  \vspace{-16pt}

  \label{fig:speech_production}
\end{figure}

\begin{itemize}
    \item We construct an interpretable benchmark for hallucination evaluation, enabling the analysis of different hallucination types.
    \item We propose the NAICL method that significantly reduces auditory hallucination rates across mainstream ALLM.
    \item We demonstrate the prevalence of auditory hallucinations in ALLMs and analyze hallucination behavior patterns.
\end{itemize}

\section{Benchmark Designing}


\subsection{Dataset Preprocessing and Filtering}

Clotho is a crowdsourced dataset for audio captioning in which each audio clip is independently annotated with natural language descriptions by five annotators. It comprises approximately 4,981 audio samples. Based on the Clotho dataset, we design and conduct a multi-stage manual filtering and revision process. First, we remove samples whose five reference captions exhibit large divergence and retain only those audio clips that show basic consistency at the event level, resulting in 3,164 samples. Each retained audio clip is then repeatedly listened to and manually verified to ensure that the events, actions, and attributes described are acoustically perceptible and consistent with the underlying acoustic evidence. In addition, subtle but reliably perceivable events that are commonly overlooked are supplemented, while any content inferred solely from human experience or scene priors without acoustic support is removed. Based on this process, we enrich the samples with event-level annotations following the AudioSet ontology, and construct the Clotho-1K evaluation dataset consisting of 1,000 audio samples. 
Each sample includes the audio, the five original human-written captions, one manually verified and revised reference caption, and AudioSet-based event annotations.

\begin{figure}[t]
  \centering 
  \includegraphics[width=\linewidth]{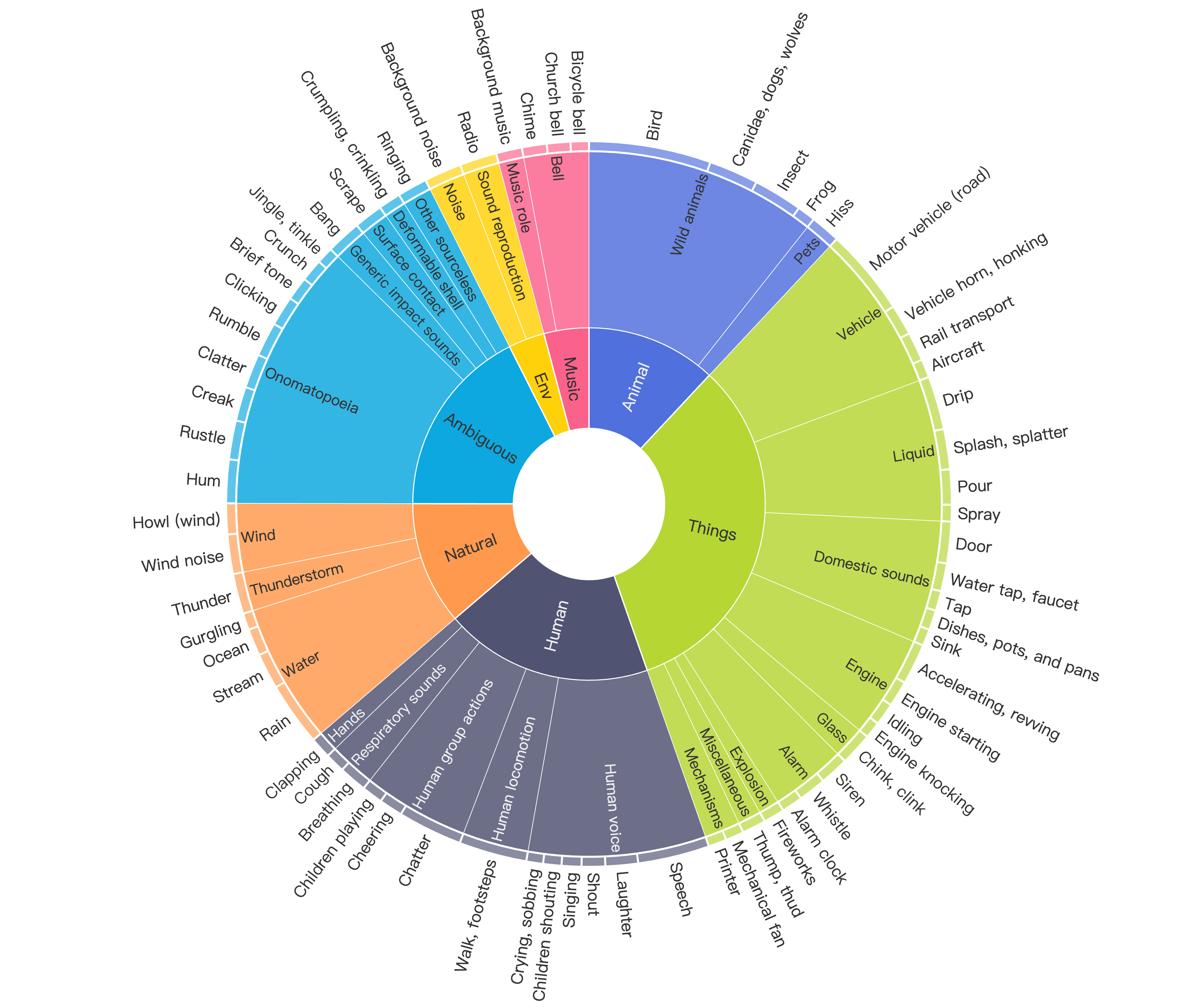}
    \caption{Hierarchy of acoustic events in Clotho-1K benchmark based on the AudioSet ontology.}  
    \vspace{-16pt}

  \label{fig:speech_production}
\end{figure}


\begin{figure*}[t]
  \centering
   \includegraphics[width=\linewidth]{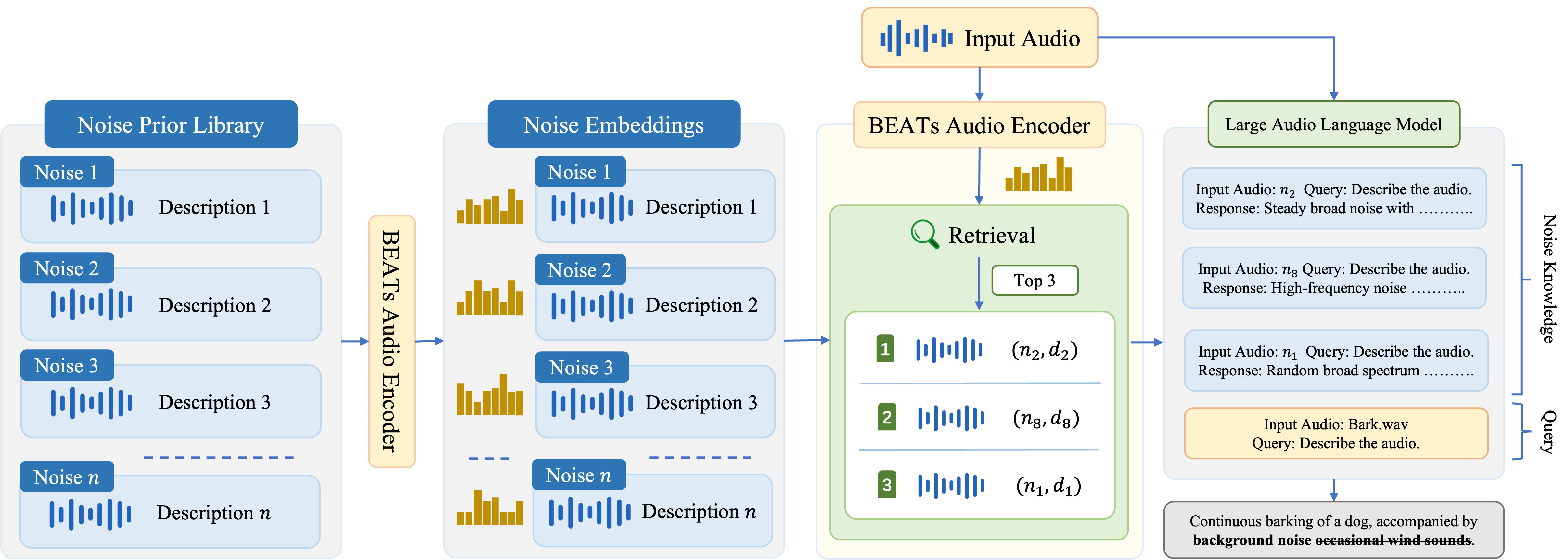}
  \caption{Workflow of NAICL, illustrating the retrieval of noise-description pairs for calibrated inference.} 
  \vspace{-16pt}
  \label{fig:speech_production}
\end{figure*}

 

\subsection{Auditory Hallucination Taxonomy}
Utilizing common patterns observed in ALLMs from audio captioning tasks as a basis, we categorize auditory hallucinations into the following four types from the perspective of generation behavior. Acoustic Attribute Hallucination: The model assigns attributes or action characteristics to events or sound sources that are indeed present in the audio, but that are not supported by acoustic evidence \cite{gunjal2024detecting}.
Source / Material Hallucination: The model incorrectly identifies the source or material of a sound, attributing an acoustic event to a sound source or object that does not correspond to the actual audio \cite{rohrbach2018object}.
Prior-driven Hallucination: The generated content is primarily driven by linguistic priors or commonsense knowledge, rather than grounded in the acoustic evidence of the input audio \cite{ji2023survey,li2023evaluating}.
Fabricated Event Hallucination: The model generates sound events that are completely absent from the input audio and have no perceptual basis at the acoustic level \cite{huang2025survey}.

\subsection{Evaluation Pipeline}

Test samples are first loaded from Clotho-1K. For each audio, the evaluated ALLM generates a description. This description is then paired with the corresponding manually revised reference description, and combined with the definitions and decision criteria of the four hallucination types to form a unified evaluation prompt. An LLM is employed as the judge (LLM-as-Judge) \cite{zheng2023judging, liu2023g} to perform detection and categorization for each sample, determining whether hallucination exists, assigning its category, and identifying the specific expressions not supported by the reference description \cite{min2023factscore}. Finally, results are aggregated across all samples to compute the HR and the occurrence ratio of each hallucination type.

\subsection{Evaluation Metrics}

The Hallucination Rate (HR) measures the frequency of erroneous model generations. It is defined as the percentage of test samples whose generated descriptions contain at least one identified hallucination type. To facilitate a more granular error analysis, we calculate the occurrence rate of each specific hallucination type across the entire test set. This metric represents the proportion of samples where a given hallucination type is present, allowing for the diagnosis of model-specific biases and weaknesses.

\section{Hallucination Mitigation Method}

In the audio modality, noise can be regarded as a lower bound of acoustic evidence. Although noise exhibits a measurable spectral structure and energy distribution, it lacks stable semantic events that can be grounded in real-world sound sources  \cite{virtanen2018computational}. When the input audio is acoustically similar to noise or exhibits high uncertainty, the model should avoid mapping such signals to specific event-level semantics. Based on this observation, we model noise as an acoustic lower-bound prior, characterizing the generation behavior that should be adopted when reliable semantic cues are insufficient.

\begin{table*}[t]
\centering
\caption{Analysis of ALLMs on Clotho-1K benchmark. HR: Hallucination Rate. Remaining columns: 4 hallucination types.}
\label{tab:hallucination_results}
\begin{tabular}{lccccc}
\toprule
\textbf{Model} &
\textbf{HR (\%)} &
\textbf{Acoustic Attribute (\%)} &
\textbf{Source (\%)} &
\textbf{Prior-Driven (\%)} &
\textbf{Fabricated (\%)} \\
\midrule
Qwen2.5-Omni-3B \cite{Xu2025QwenOmni} &
29.75 &
7.74 &
22.41 &
7.24 &
11.16 \\

StepAudio2 \cite{wu2025stepaudio2technicalreport} &
25.55 &
3.41 &
20.44 &
3.91 &
9.92 \\

Qwen-audio-chat \cite{Qwen-Audio} &
25.93 &
5.23 &
20.30 &
3.62 &
10.55 \\

Qwen2-Audio-7B-Instruct \cite{Qwen2-Audio} &
24.25 &
\textbf{3.12} &
17.61 &
7.44 &
13.68 \\

MiMo-Audio-7B-Instruct \cite{coreteam2025mimoaudio} &
26.43 &
13.27 &
19.80 &
4.12 &
10.95 \\

Kimi-Audio-7B-Instruct \cite{kimi_audio_2024} &
36.48 &
7.34 &
28.74 &
7.44 &
19.30 \\

Audio-Flamingo-3 \cite{goel2025audio} &
23.72 &
8.44 &
16.58 &
7.94 &
9.95 \\



SALMONN-7B \cite{tang2023salmonn} &
40.50 &
5.13 &
30.65 &
14.97 &
21.91 \\


Gemini-2.5-flash \cite{comanici2025gemini} &
26.85 &
13.78 &
20.26 &
4.15 &
9.93 \\

Gemini-2.5-pro \cite{comanici2025gemini} &
19.42 &
12.47 &
14.49 &
\textbf{2.31} &
\textbf{6.64} \\

GPT-audio-mini \cite{openai2025gptaudio}  &
37.69 &
23.92 &
25.53 &
4.02 &
14.59 \\

GPT-audio \cite{openai2025gptaudio}  &
21.81 &
15.28 &
15.28 &
3.02 &
7.64 \\
\midrule
Qwen2.5-Omni-7B \cite{Xu2025QwenOmni} &
26.53 &
7.24 &
18.19 &
7.84 &
10.25 \\

\textbf{Qwen2.5-Omni-7B (NAICL)} &
\textbf{16.98} &
5.63 &
\textbf{11.96} &
5.43 &
7.64 \\
\bottomrule
\end{tabular}
\end{table*}

\begin{table*}[t]
\centering
\caption{Ablation study on different configurations and noise conditions using Qwen2.5-Omni-7B on the Clotho-1K benchmark. 
HR: Hallucination Rate. A/S/P/F: 4 hallucination types. Event/Definite/Acoustic: Word-frequency metrics of the keyword sets.}

\label{tab:ablation_nic}
\begin{tabular}{lcccccccc}
\toprule
\textbf{} &
\textbf{HR (\%)} &
\textbf{A (\%)} &
\textbf{S (\%)} &
\textbf{P (\%)} &
\textbf{F (\%)} &
\textbf{Event} &
\textbf{Definite} &
\textbf{Acoustic} \\
\midrule
Qwen2.5-Omni-7B &
26.53 &
7.24 &
18.19 &
7.84 &
10.25 &
0.0831 &
0.0432 &
0.1166 \\
ICL (Real-audio) &
27.54 &
6.53 &
19.80 &
9.95 &
10.35 &
\textbf{0.0887} &
\textbf{0.0560} &
0.1340 \\
\midrule
NAICL (1-shot, w/o retrieval) &
18.99 &
7.24 &
13.07 &
7.24 &
\textbf{6.23} &
0.0714 &
0.0390 &
0.1928 \\

NAICL (2-shot, w/o retrieval) &
17.49 &
6.03 &
12.06 &
5.63 &
6.33 &
0.0656 &
0.0363 &
\textbf{0.1937} \\

NAICL (3-shot, w/o retrieval) &
17.49 &
6.33 &
11.66 &
6.33 &
5.53 &
0.0586 &
0.0350 &
0.1856 \\
\midrule

NAICL (10s, 3-shot) &
18.70 &
7.24 &
\textbf{11.46} &
\textbf{5.33} &
8.24 &
0.0563 &
0.0369 &
0.1651 \\

NAICL (Unstructured Caption) &
19.10 &
6.03 &
12.86 &
6.83 &
6.93 &
0.0739 &
0.0515 &
0.1620 \\

NAICL (2s, retrieval, 3-shot) &
\textbf{16.98} &
\textbf{5.63} &
11.96 &
5.43 &
7.64 &
0.0490 &
0.0402 &
0.1627 \\
\bottomrule
\end{tabular}
\end{table*}

Accordingly, we propose the NAICL method, as shown in Fig 3.This method operates at two complementary levels. First, it establishes lower-bound alignment in the acoustic space, reducing over-mapping from noise-like structures to specific semantic events. Second, it provides an expression template at the generation level, encouraging reduced semantic commitment when evidence is insufficient. We construct a structured noise prior library $\mathcal{L}$ consisting of diverse broadband noise samples paired with conservative textual descriptions. The library covers different spectral and energy distributions, and each noise sample is associated with a structurally consistent conservative description. These descriptions primarily employ acoustics-level expressions and avoid concrete event verbs or explicit entity assertions (e.g., using ``continuous background noise'' or ``irregular low-frequency sound'' instead of specific sound sources). Such descriptions serve as behavioral templates, encouraging the model to adopt abstract acoustics-level expressions under insufficient evidence rather than making deterministic event-level predictions.

This study implements a retrieval module. Given an input audio $a$, the system retrieves the $K$ most acoustically similar noise-description pairs from the library $\mathcal{L}$ based on embedding similarity. We employ BEATs \cite{chen2022beats} as the acoustic encoder to extract high-level representations from both the input audio and the noise library. The retrieved set is defined as

\begin{equation}
C_{\text{noise}} = \text{Retrieve}(a, \mathcal{L}, K) 
= \{(n_k, d_k)\}_{k=1}^{K},
\end{equation}

\noindent where $n_k$ denotes the $k$-th retrieved noise sample and $d_k$ its corresponding conservative description.

During inference, the ALLM $M$ is conditioned on both the input audio and the retrieved noise context. The generated caption $\hat{y}$ is obtained as

\begin{equation}
\hat{y} = \arg\max_{y} p_M(y \mid a, C_{\text{noise}}),
\end{equation}

\noindent where $y$ represents a candidate caption sequence, and $p_M(y \mid a, C_{\text{noise}})$ is the conditional probability assigned by $M$ given the input audio $a$ and the retrieved noise context $C_{\text{noise}}$.

We further examine several NAICL design variants, including different noise durations, varying shot numbers, the use or removal of the retrieval module, and structured versus unstructured description formats, and compare them with real-audio in-context learning (ICL). We constructed three keyword sets corresponding to different levels of semantic commitment,as follows. Event Verbs: capturing whether the model generates concrete and causally attributable acoustic events. Definite Terms: capturing whether the model makes explicit assertions about entities, attributes, or temporal states. Acoustic Terms: capturing whether the model adopts more conservative, acoustics-level descriptions. Let the test set contain $N$ samples, and let $y_i$ denote the generated caption for the $i$-th sample. To quantify the occurrence of various hallucination-related descriptors, we define the sample-level frequency for each keyword category $k \in \{\textit{event}, \textit{definite}, \textit{acoustic}\}$ as

\begin{equation}
\mathrm{Freq}_k = \frac{1}{N} \sum_{i=1}^{N} \mathbb{I}(\exists w \in V_k, w \in y_i),
\end{equation}

\noindent where $\mathbb{I}(\cdot)$ is the indicator function that outputs 1 if the generated sequence $y_i$ contains at least one keyword $w$ from the predefined set $V_k$, and 0 otherwise.
 
\section{Results and Discussion}

\subsection{Experiment Setup}
The retrieval module adopts the officially fine-tuned BEATs model as the acoustic encoder. Retrieval is performed using cosine similarity in the embedding space, dynamically selecting the Top-3 most relevant noise–description pairs from the structured noise prior library for each input. All noise samples are synthetically generated and paired with descriptions following a unified structural template; the duration of each noise segment is fixed at 2 seconds. NAICL is primarily implemented and tested on Qwen2.5-Omni-7B.  For hallucination evaluation, we employ Qwen3-Next-80B-A3B-Instruct \cite{qwen3technicalreport,qwen2.5-1m} as the LLM-as-Judge to perform automatic detection and categorization of hallucination types. In addition, we construct three keyword sets corresponding to different levels of semantic commitment (Event, Definite, Acoustic), each containing 30 terms. More detailed implementation configurations and parameter settings are publicly available in our GitHub repository \url{https://github.com/OrgHuang/NAICL-Clotho1k.git}.

\subsection{Benchmark Evaluation and Mitigation Results}
As shown in Table~\ref{tab:hallucination_results}, all evaluated models exhibit substantial hallucination rates. Among the categories, \textit{Source Material} and \textit{Fabricated Event} are the dominant types, while \textit{Prior-Driven} hallucination appears consistently across models.  We further observe that GPT-audio-mini, Kimi-Audio-7B and SALOMNN-7B suffer from severe hallucination problems. We further find that GPT-audio-mini tends to produce predominantly acoustics-level abstract descriptions with weak event expressiveness, whereas Kimi-Audio generates more concrete and detailed content. In contrast, multimodal large language models (MLLMs) \cite{Xu2025QwenOmni,comanici2025gemini} exhibit a more stable performance. When using the NAICL, the overall hallucination rate decreases from 26.53\% to 16.98\%, with consistent reductions across all types. These findings demonstrate that certain audio clips may admit multiple plausible event interpretations, while paired supervision during training enforces deterministic mappings, which naturally gives rise to attribute and object-level hallucinations.

\subsection{Ablation Study for NAICL}
As shown in Table~2, using real audio as few-shot examples does not yield effective improvements. This suggests that real audio demonstrations often carry strong semantic patterns and scene priors, which tend to further reinforce event-level deterministic expressions. According to keyword frequency statistics, when fixed noise examples are utilized without the retrieval module, the hallucination rate decreases; however, the model exhibits a clear tendency toward over-conservative generation. In the comparison of noise duration, 2-s noise performs better than 10-s noise, since longer noise segments introduce excessive contextual length and additional interference. Moreover, structured noise descriptions outperform unstructured ones \cite{cho2025mechanism}, thereby enhancing the constraint and stability of ICL on generation behavior. Finally, frequency analysis shows that after introducing NAICL, the occurrence of \textit{Event} and \textit{Definite} terms decreases significantly, while the frequency of \textit{Acoustic} terms increases. This shift indicates that the model adopts more conservative and acoustically grounded expressions when describing ambiguous events, thereby explaining the reduction in hallucination rates.

\section{Conclusion}
In this study, we proposed the NAICL method and constructed a Clotho-1K benchmark to evaluate hallucinations in ALLMs. The proposed treats noise as an acoustic lower-bound prior and regulates semantic commitment under insufficient evidence through structured noise examples. The results demonstrate that NAICL significantly suppresses hallucinations, providing an effective inference-time calibration scheme to enhance the reliability of ALLMs in complex acoustic environments. Future work will expand the dataset scale and increase scenario diversity to improve evaluation coverage and explore fine-tuning methods better suited for audio in order to further enhance semantic calibration under acoustically complex conditions.

\bibliographystyle{IEEEtran}
\bibliography{mybib}

\end{document}